\documentclass[sigconf, nonacm]{acmart}

\AtBeginDocument{%
  }

\usepackage{lipsum}
\usepackage{booktabs} 
\usepackage{url} 
\usepackage{caption}
\usepackage{pdfpages}
\usepackage{multirow}
\usepackage{kotex}
\usepackage{subcaption} 
\usepackage{listings}
\usepackage{xcolor}
\usepackage{subcaption}
\usepackage{adjustbox} 
\usepackage{subcaption}
\usepackage{enumitem}
\usepackage{dirtree}
\usepackage{mdframed}

\definecolor{codebg}{RGB}{245,245,244}
\definecolor{keyword}{RGB}{0,0,180}
\definecolor{comment}{RGB}{0,128,0}
\definecolor{string}{RGB}{163,21,21}

\lstdefinestyle{pythonstyle}{
  language=Python,
  keywordstyle=\color{blue},
  stringstyle=\color{red},
  commentstyle=\color{gray},
  numbers=left,
  numberstyle=\tiny,
  stepnumber=1,
  frame=single,
  breaklines=true,
  xleftmargin=1em,   
  xrightmargin=1em, 
  framexrightmargin=0pt 
}

\lstdefinestyle{bashstyle}{
  language=bash,
  backgroundcolor=\color{codebg},
  keywordstyle=\bfseries\color{keyword},
  commentstyle=\itshape\color{comment},
  stringstyle=\color{string},
  frame=single,
  framesep=6pt,
  frameround=tttt,
  numberstyle=\tiny\color{gray},
  stepnumber=1,
  showstringspaces=false,
  tabsize=2,
  breaklines=true,
  breakatwhitespace=true,
  captionpos=b,
  aboveskip=6pt, 
  belowskip=6pt
}

\begin{document}

\title[Analyticup E-commerce Product Search Competition]{%
  Analyticup E-commerce Product Search Competition \\
  {Technical Report from Team Tredence\_AICOE}}

\author{Rakshith R}
\thanks{
Presented at Analyticup Workshop, held in conjunction with the 34th ACM International Conference on Information and Knowledge Management (CIKM’25), November 10–14, 2025, Seoul, Republic of Korea.
}
\affiliation{%
  \institution{Tredence}
  \country{India}
}
\email{rakshith.r@tredence.com}

\author{Shubham Sharma}
\affiliation{%
  \institution{Tredence}
  \country{India}}
\email{shubham.sharma@tredence.com}

\author{Mohammed Sameer Khan}
\affiliation{%
  \institution{Tredence}
  \country{India}}
\email{mohammed.sameerkhan@tredence.com}

\author{Ankush Chopra}
\affiliation{%
  \institution{Tredence}
  \country{India}}
\email{ankush.chopra@tredence.com}

\renewcommand{\shortauthors}{Team Tredence\_AICOE}

\begin{abstract}
    This study presents the multilingual e-commerce search system
    developed by the Tredence\_AICOE team. The competition features two multilingual relevance tasks: Query-Category (QC) Relevance, which evaluates how well a user’s search query aligns with a product category, and Query-Item (QI) Relevance, which measures the match between a multilingual search query and an individual product listing. To ensure full language coverage, we performed data augmentation by translating existing datasets into languages missing from the development set, enabling training across all target languages. We fine-tuned Gemma-3 12B and Qwen-2.5 14B model for both tasks using multiple strategies. The Gemma-3 12B (4-bit) model achieved the best QC performance using original and translated data, and the best QI performance using original, translated, and minority class data creation. These approaches secured 4th place on the final leaderboard, with an average F1-score of 0.8857 on the private test set.
\end{abstract}




\maketitle

\section{Introduction}

In recent years, e-commerce platforms have increasingly leveraged multilingual search capabilities to deliver relevant results across diverse markets. The CIKM AnalytiCup 2025 competition (Multilingual Query–Item \& Query–Category Relevance for E-commerce Search) invited participants to build systems that accurately model relevance in two tasks. 

The first task Query-to-Category (QC) required systems to verify whether a hierarchical category mapping was appropriate for a given user query. The second task Query-to-Item (QI) focused on determining whether an individual product listing accurately matched the search query. Both tasks demanded robust multilingual modelling, high precision under data imbalance, and adaptability to real-world e-commerce noise. 

Our team, Tredence\_AICOE, approached these challenges by combining data augmentation, multilingual fine-tuning, and targeted negative sampling strategies within the competition’s computational limits (models restricted to less than 15 billion parameters). Specifically, we expanded language coverage by translating queries from existing languages into missing ones, addressing skewed class distributions by converting positive samples into negative samples through category and item perturbations. We fine-tuned Gemma-3 12B \cite{gemmateam2025gemma3technicalreport} and Qwen-3 14B \cite{qwen3} models, observing that Gemma consistently outperformed Qwen across all experiments. 

Through this approach, our system achieved strong multilingual generalization, ranking 4th on the final competition leaderboard. The following sections discuss the key experimental findings, reflections, and lessons learned from our participation in this multilingual relevance challenge. 
\section{Dataset Statistics}
\subsection{Query-Category Path}

\begin{figure}[h]
  \includegraphics[width=\columnwidth]{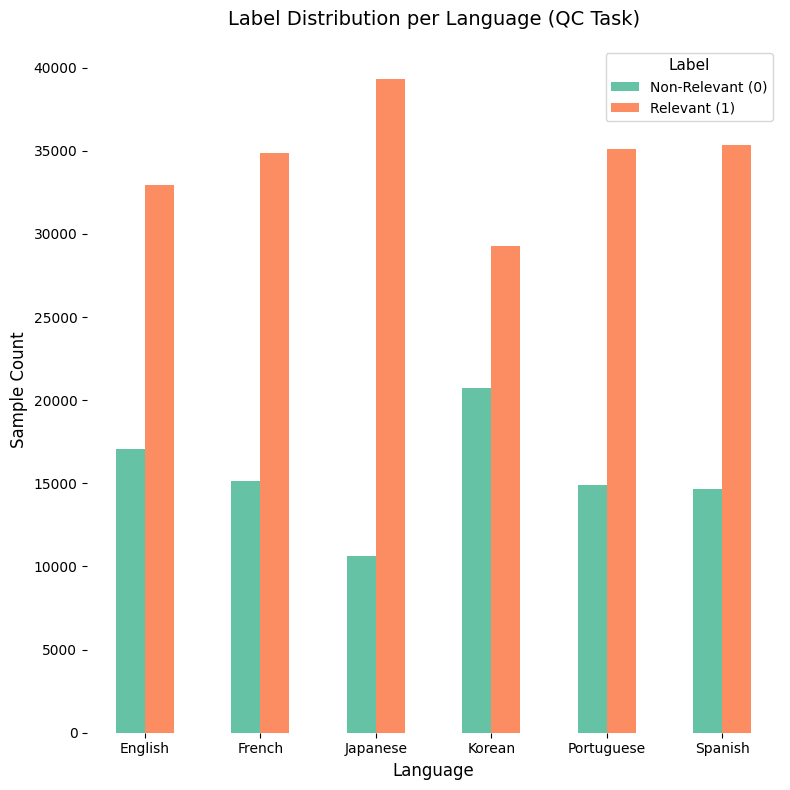}
  \caption{Language-wise distribution of positive and negative samples in the QC training dataset before data augmentation.}
  \label{fig:ld_before}
\end{figure}

The QC training dataset comprised of 300,000 samples across six languages - English, French, Korean, Spanish, Portugese, and Japanese. The developement set included ten languages in total, adding German, Italian, Polish, and Arabic, for which no training data was available. Each sample included a search query and a category path, where the query appeared in the respective language, but the category path remained in English across all languages. For example, a query such as “wireless headphones” might correspond to the category path “Electronics, Audio, Headphones”. In this path, L0 represents Electronics, L1 denotes Audio, and L2 indicates Headphones. The training data was highly imbalanced, with 206,852 positive samples and 93,148 negative samples. A detailed breakdown of the positive and negative samples across languages is presented in Figure~\ref{fig:ld_before}.

\subsection{Query-Item}

\begin{figure}[h]
  \includegraphics[width=\columnwidth]{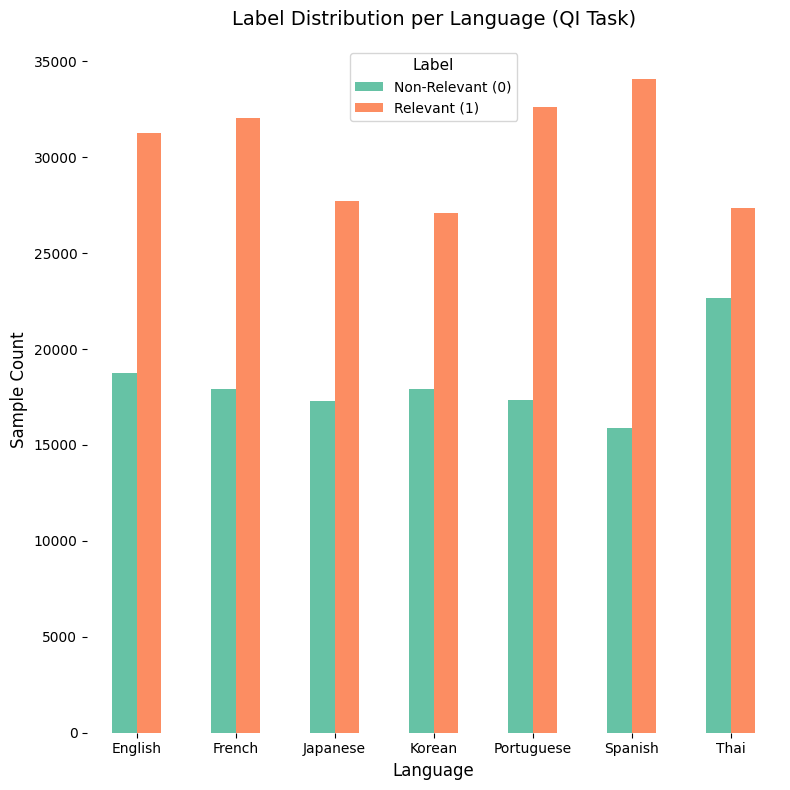}
  \caption{Language-wise distribution of positive and negative samples in the QI training dataset before data augmentation.}
  \label{fig:LC_QI}
\end{figure}

The QI training dataset comprised of 340,000 samples across  English, Spanish,  Portuguese, Thai, Japanese, Korean. The development set included thirteen languages in total, adding German, Italian, Polish, Arabic, Vietnamese, Indonesian	 for which no training data was available. Each sample included search query and an item , where the query appeared in the respective languages but the item title appeared in English across all languages. The training data was highly imbalanced with 212,199 positive samples and 127,801 negative samples across languages is present in Figure~\ref{fig:LC_QI}.

\section{Data Preparation and Augmentation}
\subsection{Query-Category Path}
To address the gap in language coverage, additional training data was generated for the languages present in the development set but missing from the training data - German, Arabic, Italian, and Polish. Queries from existing languages were translated into the target languages while keeping the category path unchanged, maintaining alignment with the original data structure. Translations were performed using Gemini-2.5 Flash \cite{comanici2025gemini25pushingfrontier}.

To ensure balanced coverage, datapoints were sampled from subsets whose category paths matched those found in the development set. This process produced approximately 42,000 translated samples for each missing language, expanding the dataset’s linguistic diversity. In addition, incorrect language tags in the provided data were corrected using Gemini 2.5 Flash \cite{comanici2025gemini25pushingfrontier}, though no experiments were conducted using these corrected-tag samples.

\begin{figure}[h]
  \includegraphics[width=\columnwidth]{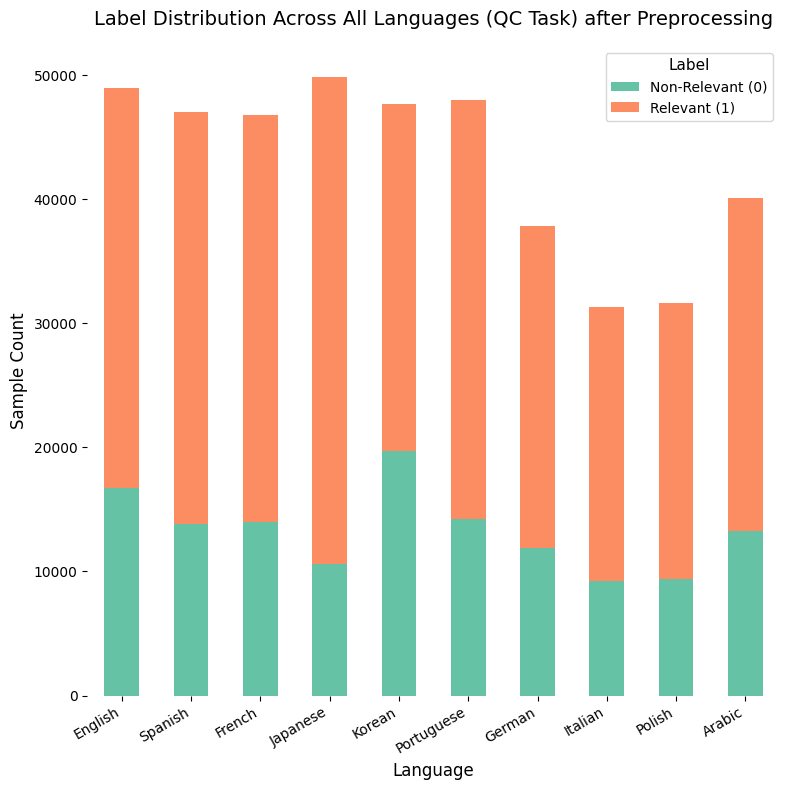}
  \caption{Language-wise distribution of positive and negative samples in the QC training dataset after data augmentation.}
  \label{fig:ld_after}
\end{figure}

As part of data preparation, we removed records with conflicting labels for the same query–category pair, deduplicated identical entries, and filtered purely numerical queries, retaining only those with meaningful search intent verified through manual review. A detailed breakdown of the positive and negative samples after data augmentation across languages is presented in Figure~\ref{fig:ld_after}.

\subsection{Query-Item}
The training data for the Query–Item (QI) relevance model exhibited a linguistic coverage gap, as several languages represented in the development set—specifically German, Arabic, Italian, Polish, Vietnamese, and Indonesian—were absent from the training corpus. To address this discrepancy, the query texts were translated into the underrepresented languages, while the item titles were retained in their original form to preserve semantic consistency. Translations were generated using Gemini 2.5 Flash \cite{comanici2025gemini25pushingfrontier}, wherein English-language queries were systematically translated into each of the six target languages, yielding approximately 50,000 synthetic samples per language. This data augmentation strategy was designed to enhance multilingual robustness and facilitate improved cross-lingual generalization within the QI relevance model.

\begin{figure}[h]
  \includegraphics[width=\columnwidth]{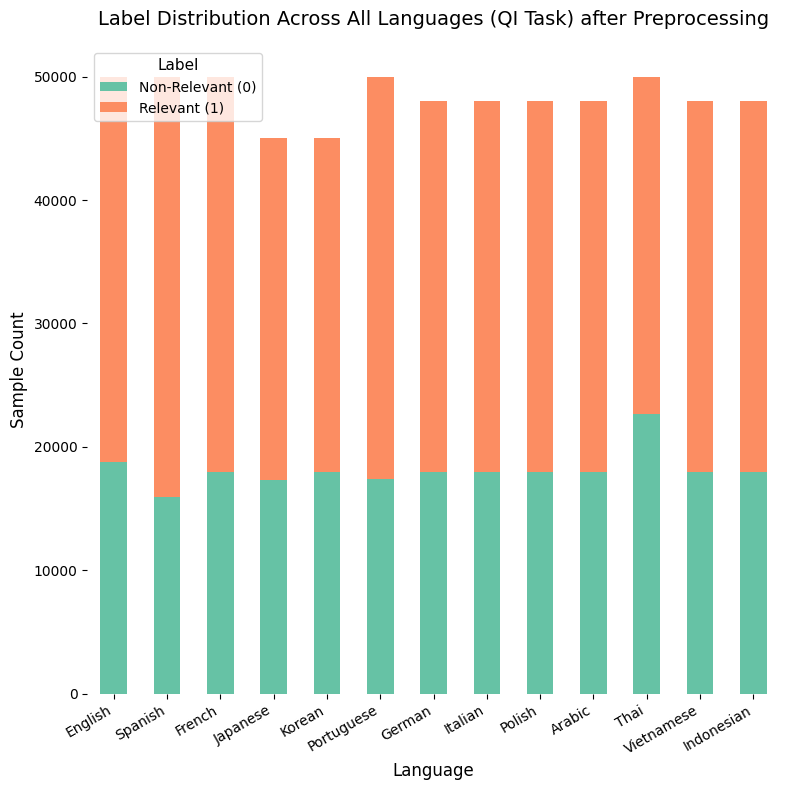}
  \caption{Language-wise distribution of positive and negative samples in the QI training dataset after data augmentation.}
  \label{fig:QI_after}
\end{figure}

To ensure the reliability and consistency of the dataset, a series of preprocessing steps were applied. Records exhibiting label inconsistencies, where identical query–item pairs were assigned conflicting relevance labels, were removed to maintain unique and unambiguous annotations. Duplicate entries containing identical combinations of query, item, and label were also eliminated, ensuring that each instance was represented only once while preserving data integrity. This deduplication process reduced redundancy without any loss of information. A detailed breakdown of positive and negative samples after data augmentation across languages is presented in Figure~\ref{fig:QI_after}.

\section{Experimentations}

\begin{table*}[htbp]
  \centering
  \caption{Performance of fine-tuning experiments on the Query–Category (QC) task during the preliminary and final rounds. }
  \renewcommand{\arraystretch}{1.2}
  \setlength{\tabcolsep}{8pt}
  \begin{tabular}{lccc}
    \toprule
    \textbf{Experiment} & \textbf{Model Used} & \textbf{Dev Set Score} & \textbf{Private Test Set Score} \\
    \midrule
    E1 & Gemma-3 12B (4b) & \textbf{0.8947} & \textbf{0.8930} \\
    E2 & Gemma-3 12B (4b) & 0.8930 & -- \\
    E3 & Gemma-3 12B (4b) & 0.8909 & -- \\
    E4 & Gemma-3 12B (4b) & 0.8792 & -- \\
    E5 & Gemma-3 12B (4b) & -- & 0.8745 \\
    E5 & Qwen-3 14B (4b) & -- & 0.8682 \\
    \bottomrule
  \end{tabular}
  \label{tab:qc_experiments_full}
\end{table*}

\begin{table*}[htbp]
  \centering
  \caption{Performance of fine-tuning experiments on the Query–Item (QI) task during the preliminary and final rounds.}
  \renewcommand{\arraystretch}{1.2}
  \setlength{\tabcolsep}{8pt}
  \begin{tabular}{lccc}
    \toprule
    \textbf{Experiment No.} & \textbf{Model Used} & \textbf{Dev Set Score} & \textbf{Private Test Set Score} \\
    \midrule
    E1 & Gemma-3 12B (4b) & 0.8792 & -- \\
    E2 & Gemma-3 12B (4b) & \textbf{0.8804} & \textbf{0.8779} \\
    E2 & Gemma-3 12B (16b) & -- & 0.8765 \\
    E3 & Gemma-3 12B (4b) & 0.8798 & -- \\
    \bottomrule
  \end{tabular}
  \label{tab:qi_experiments_full}
\end{table*}

\subsection{Query-Category Path}

We conducted a series of fine-tuning experiments to evaluate the effect of multilingual data augmentation and negative sampling strategies on model performance. Each experiment was fine-tuned using a zero-shot prompt for five epochs with early stopping and a learning rate of 2e-4. We performed LoRA-based fine-tuning using the Unsloth framework \cite{unsloth} with a rank of 32. 

In our first experiment (E1), we performed a straightforward fine-tuning on the combined dataset containing both the original six languages and the four translated languages. Both Gemma-3 12B (4-bit) \cite{gemmateam2025gemma3technicalreport} was fine-tuned under this setup. 

In our second experiment (E2), we focused on addressing label imbalance by introducing strong negative samples. For each positive query–category pair, we retained the same Level-1 (L1) category but replaced the Level-4 (L4) category with another valid L4 under the same hierarchy (e.g., sneakers - hiking shoes). This approach introduced harder negatives within semantically similar category branches. Gemma-3 12B (4-bit) \cite{gemmateam2025gemma3technicalreport} model was finetuned on this setup. 

In our third experiment (E3), we extended this idea by creating negative samples that shared all but the final level of the category path. This ensured that negatives were contextually similar to positives while differing at the finest level of categorization. Gemma-3 12B (4-bit) \cite{gemmateam2025gemma3technicalreport} model was finetuned on this setup.

In our fourth experiment (E4), we altered the data splitting strategy to ensure query uniqueness across all splits. No query appeared in more than one of the training, validation, or test sets, effectively preventing data leakage and improving evaluation robustness. Gemma-3 12B (4-bit) \cite{gemmateam2025gemma3technicalreport} model was finetuned on this setup.

In out final experiment (E5), we generated synthetic negative queries for originally positive query–category pairs, keeping the category path unchanged. These negative queries were generated using Gemma-3 27B \cite{gemmateam2025gemma3technicalreport}, and fine-tuning was carried out with Gemma-3 12B (4-bit) \cite{gemmateam2025gemma3technicalreport} and Qwen-3 14B (4-bit) \cite{qwen3}. This experiment was conducted after the preliminary round and evaluated exclusively on the final-round test set.

The first experiment (E1), achieved the best performance on the development set during the preliminary round. The best model weights from this setup were used for inference on the private test set, as none of the other experiments outperformed E1 on the dev set. To further validate this setup, we performed a full-precision fine-tuning of the same model. However, the results were inferior to the 4-bit configuration. A summary of F1-scores across all experiments on the preliminary dev set and the final private test set is shown in Table~\ref{tab:qc_experiments_full}.

\subsection{Query-Item}

We conducted a series of fine-tuning experiments to evaluate the effect of multilingual data augmentation and negative sampling strategies on model performance. We performed LoRA-based fine-tuning using the Unsloth framework \cite{unsloth} for five epochs with early stopping and a learning rate of 2e-4. The model was trained as a binary classifier with a rank of 32. 

In our first experiment (E1), we performed a straightforward fine-tuning on the combined dataset containing both the original seven languages and the six translated languages. The underlying hypothesis was that increasing language diversity would enable the model to capture broader linguistic variations. Gemma-3 12B (4-bit) \cite{gemmateam2025gemma3technicalreport} was finetuned under this setup.  

In our second experiment (E2), we incorporated easy negative samples to further refine the model’s discriminative capabilities. These samples were constructed using multilingual embeddings for French, Spanish, Japanese, Korean, and Portuguese. For each positive query–item pair, candidate items from the same language (excluding the correct item) are identified. Cosine similarity between the positive item and each candidate is computed using normalized embeddings. The candidate with the lowest similarity score is chosen as the easy negative. Gemma-3 12B (4-bit) \cite{gemmateam2025gemma3technicalreport} was finetuned under this setup.

In our third experiment (E3), the focus shifted toward hard negative sampling to enhance the model’s sensitivity to subtle semantic differences. Using the same set of five auxiliary languages — French, Spanish, Japanese, Korean, and Portuguese — the goal was to help the model better distinguish between items that appear related but are not exact matches. For each correct query–item pair, other items from the same language were compared using cosine similarity with the positive item. Candidates with a similarity score below 0.7 were considered potential hard negatives, representing examples that were somewhat similar yet still incorrect. Unlike easy negatives, these samples contained partial semantic overlap, making them more challenging for the model to differentiate. Gemma-3 12B (4-bit) \cite{gemmateam2025gemma3technicalreport} was then finetuned under this setup.

The second experiment (E2), achieved the best performance on the development set during the preliminary round. The best model weights from this setup were used for inference on the private test set, as none of the other experiments outperformed E2 on the dev set. To further validate this setup, we performed a full-precision fine-tuning of the same model. However, the results were inferior to the 4-bit configuration. A summary of F1-scores across all experiments on the preliminary dev set and the final private test set is shown in Table~\ref{tab:qi_experiments_full}.

\section{Discussion}
Both tasks faced challenges of incomplete multilingual coverage and label imbalance. We mitigated these by expanding the datasets through targeted translations and thorough data cleaning to ensure balanced linguistic representation and consistent labeling quality.Model fine-tuning began with Gemma-3 12B (4-bit) \cite{gemmateam2025gemma3technicalreport} due to it’s multilingual capabilities and the 15B model limit set by the organizers. Error analysis showed a tendency to overpredict positives.

To mitigate this for the Query-Category Path (QC) task, we explored various negative-generation strategies by perturbing category hierarchies. Changing the final node in category paths proved intuitive but occasionally could have introduced noise when the replacement happened to be a semantically close and valid category. Restricting replacements to categories outside the same hierarchy produced little change, suggesting the original possible noise was not a contributing factor. We also tested splitting data such that no query appeared in multiple subsets, but this had minimal effect on model generalization. 

A further strategy involved generating synthetic negative queries for existing positive examples using a larger model. While this helped address the imbalance more directly, the generated text lacked the linguistic consistency of natural data, leading to a slight drop in performance. Ultimately, the most stable and generalizable configuration remained the version trained on the combination of original and translated data, highlighting that controlled multilingual expansion was more beneficial than aggressive balancing. 

For the Query-Item (QI) task, we followed a similar multilingual augmentation approach, translating English queries into six missing languages to better align with the evaluation set. The performance gain was smaller compared to QC, likely due to using only English as the translation source, which limited linguistic diversity. We then experimented with constructing negative samples based on semantic similarity: pairing queries with obviously irrelevant (easy) or subtly similar (hard) items. The easy negatives improved precision and overall stability, whereas the hard negatives increased misclassifications because of ambiguous near-matches. 

Across both tasks, a few consistent insights emerged. Translation quality mattered more than quantity, high-fidelity multilingual data yielded stronger cross-lingual generalization than larger but noisier synthetic corpora. Moderate class imbalance proved tolerable, as fine-tuned LLMs inherently captured relevance asymmetries without requiring strict label balance. Finally, semantic coherence, ensuring that augmented or modified pairs remained logically valid was critical to maintaining stable learning. In retrospect, greater emphasis on improving the diversity and quality of data augmentation might have yielded stronger gains, as much of our later experimentation was focused on mitigating class imbalance rather than enriching the underlying data itself. 
\section{Conclusion}
This work presented our approach and learnings from the CIKM AnalytiCup 2025 multilingual e-commerce search challenge, where our team ranked fourth overall. We addressed two binary relevance tasks Query-to-Category and Query-to-Item under constraints of limited language coverage, label imbalance, and model size. Our strategy centered on multilingual data augmentation, label balancing through negative sample creation and efficient fine-tuning of large language models, primarily Gemma-3 12B \cite{gemmateam2025gemma3technicalreport}, which consistently outperformed Qwen-3 14B \cite{qwen3} across experiments. 

By translating queries into missing languages and performing targeted label balancing through negative sample creation, we built training datasets that were both linguistically richer and more representative of real-world distributions. We conducted several experiments to address the data imbalance by creating more class 0 data in several ways. These techniques proved particularly effective in the Query-Item task, where easy-negative sampling improved generalization, while the Query-Category task benefited most from multilingual expansion rather than further synthetic balancing. 

Overall, our findings highlight that for multilingual relevance modelling, high-quality translation-based augmentation and controlled fine-tuning yield more reliable improvements than heavy rebalancing or aggressive data manipulation.


\bibliographystyle{ACM-Reference-Format}
\bibliography{reference}

\end{document}